\documentclass[prd, amsmath, floats, floatfix, twocolumn, nopreprintnumbers, superscriptaddress, nofootinbib, showpacs, eqsecnum]{revtex4-1}

\usepackage{ulem}

\usepackage{color}
\usepackage{graphicx}
\usepackage{float}
\usepackage{amsmath}
\usepackage{acronym}
\usepackage{multirow}
\usepackage{epsfig}
\usepackage{verbatim}
\usepackage[squaren]{SIunits}
\addunit{\Mparsec}{Mpc}
\addunit{\smass}{M_{\odot}}

\newcommand{\bea}{\begin{eqnarray}} 
\newcommand{\eea}{\end{eqnarray}}

\newcommand{\beq}{\begin{equation}} 
\newcommand{\eeq}{\end{equation}} 

\begin{document}

\title{Towards Rapid Parameter Estimation on Gravitational Waves from Compact Binaries using Interpolated Waveforms}

\author{R. J. E. Smith}
\affiliation{School of Physics and Astronomy, University of Birmingham, Edgbaston, Birmingham B15 2TT, UK}
\affiliation{Perimeter Institute for Theoretical Physics, Waterloo, Ontario N2L 2Y5, Canada}

\author{K. Cannon}
\affiliation{Canadian Institute for Theoretical Astrophysics, 60 St.\ George St., Toronto, ON, M5S 3H8, Canada}

\author{C. Hanna}
\affiliation{Perimeter Institute for Theoretical Physics, Waterloo, Ontario N2L 2Y5, Canada}

\author{D. Keppel}
\affiliation{Albert-Einstein-Institut, Max-Planck-Institut f\"{u}r Gravitationsphysik, D-30167 Hannover, Germany}
\affiliation{Leibniz Universit\"{a}t Hannover, D-30167 Hannover, Germany}

\author{I. Mandel}
\affiliation{School of Physics and Astronomy, University of Birmingham, Edgbaston, Birmingham B15 2TT, UK}

\begin{abstract}
Accurate parameter estimation of gravitational waves from coalescing compact binary sources is a key requirement for gravitational-wave astronomy. Evaluating the posterior probability density function of the binary's parameters (component masses, sky location, distance, etc.) requires computing millions of waveforms.  The computational expense of parameter estimation is dominated by waveform generation and scales linearly with the waveform computational cost. Previous work showed that gravitational waveforms from non-spinning compact binary sources are amenable to a truncated singular value decomposition, which allows them to be reconstructed via interpolation at fixed computational cost.  However, the accuracy requirement for parameter estimation is typically higher than for searches, so it is crucial to ascertain that interpolation does not lead to significant errors.  Here we provide a proof of principle to show that interpolated waveforms can be used to recover posterior probability density functions with negligible loss in accuracy with respect to non-interpolated waveforms. This technique has the potential to significantly increase the efficiency of parameter estimation.
\end{abstract}

\maketitle
\section{Introduction}

Astronomy and tests of fundamental physics with gravitational waves from compact binary coalescence (CBC) will ultimately be limited by our ability to estimate the binary's source parameters from the gravitational-wave signal \cite[e.g.,][]{Li:2012, MandelO:2010, BB:2003}. CBC sources with total masses in the range $\unit{2}{\smass} \lesssim M_{\mathrm{T}} \lesssim \unit{500}{\smass}$ will be amongst the prime sources for advanced LIGO \cite{Harry:2010} and VIRGO \cite{aVirgo} when they begin operating around 2015 \cite{LSCVrates:2010}.

In a Bayesian treatment of parameter estimation, one is interested in the probability distribution of the set of source parameters of the underlying model given an observed stretch of data.  
Waveform computation represents the majority of the computation cost in the Bayesian analysis of CBC sources, so the total computational cost scales roughly linearly with waveform generation. This becomes burdensome when one needs to explore a large dimensional parameter space as the number of waveform computations is large, e.g., $\mathcal{O}(10^{6})$ \cite{Raymond:2009}.

Recently, Cannon \textit{et al.}~\cite{Cannon:2010} showed that a truncated singular value decomposition (SVD) can be applied to gravitational-wave template banks which span the two parameters describing the masses of the coalescing binary. The SVD decomposes the bank into a set of ``basis templates'' and projection coefficients. In general, the number of basis templates is much less than the total number of templates in the bank. Furthermore, the projection coefficients can be interpolated across the domain of the bank \cite{Cannon:2012}. Template waveforms can thus be interpolated. Because the \textit{intrinsic} parameter space of CBC sources with non-spinning components is two-dimensional (the two mass parameters), it is possible to set up the waveform computation for parameter estimation such that the waveform calculations are done by interpolation alone.  However, the errors incurred from interpolation could, in principle, affect parameter-estimation accuracy.

In this paper, we describe the application of SVD-interpolated waveforms to CBC parameter estimation.  For a simulated data set containing a gravitational wave signal we provide a proof of principle that SVD-interpolated waveforms can be used for parameter estimation without significantly affecting the accuracy of the inferred probability distributions of the source parameters. We further show that the computational cost of using interpolated waveforms is around an order of magnitude less than that of commonly-used time-domain waveform families. This technique has the potential to increase the computational efficiency of CBC parameter estimation when the computational cost is dominated by waveform computation. Our application of the SVD is limited to a small patch of parameter space about the injected signal value.

This paper is organized as follows. In Sections~\ref{sec:cbc_est} and \ref{sec:interp_svd} we outline the principles of parameter estimation for CBCs and interpolating template waveforms based on the SVD, respectively. In Sec.~\ref{sec:param_est} we describe the application of SVD-interpolated waveforms to parameter estimation. In Sec.~\ref{sec:results} we compare the results of using interpolated and non-interpolated waveforms for parameter estimation and compare the computational cost of interpolation to using non-interpolated waveform families. In Sec.~\ref{sec:conc} we consider the future of using SVD-interpolated waveforms for parameter estimation and discuss the technical requirements of implementing these waveforms in parameter-estimation pipelines.

\section{CBC Parameter Estimation with MCMC}
\label{sec:cbc_est}
The central quantity of interest in Bayesian parameter estimation is the posterior probability density function (PDF) of a set of parameters $\vec{\theta}$ which parameterize a model, $\mathcal{H}$, assumed to  describe a data set $d$. The PDF is related to the likelihood function and prior probability via Bayes' theorem and is given by
\beq
p(\vec{\theta}| d, \mathcal{H}) = \frac{\mathcal{P}( \vec{\theta} | \mathcal{H} )\ \mathcal{L}(d | \vec{\theta}, \mathcal{H} ) } { p(d | \mathcal{H} ) },\
\label{eq:bayes}
 \eeq
 where $\mathcal{L}(d | \vec{\theta}, \mathcal{H} )$ is the likelihood function and $\mathcal{P}( \vec{\theta} | \mathcal{H} )$ is the prior probability which encodes our \textit{a priori} belief in the distribution of $\vec{\theta}$. The quantity in the denominator, $p(d | \mathcal{H} )$, is known as the evidence and is an overall normalization factor which we will not deal with here. 
 
The CBC parameter vector $\vec{\theta}$ is high-dimensional. The phasing and amplitude of a waveform from a non-spinning coalescing compact binary source is controlled by two mass parameters, the chirp mass $\mathcal{M} = (m_{1}m_{2})^{3/5}/(m_{1} + m_{2})^{1/5}$ and symmetric mass ratio $\eta = (m_{1}m_{2})/(m_{1} + m_{2})^{2}$,  where $m_{1}$ and $m_{2}$ are the component masses of the binary. In addition, a gravitational wave source with respect to the earth is specified by location dependent parameters. These are the distance from the Earth $D$, inclination $\iota$, right ascension $\alpha$, declination $\delta$, polarization phase $\varphi$ and time and phase at coalescence, $t_{c}$ and $\phi_{c}$. In general, the CBC parameter vector $\vec{\theta}$ is nine-dimensional for circular binaries with non-spinning components.

One of the goals of gravitational-wave astronomy is to estimate the PDF of the parameters of a candidate gravitational wave source in order to assign a meaningful probability to our measurements of the source properties and demographics. To compute the right hand side of \eqref{eq:bayes}, we directly evaluate the likelihood function, $\mathcal{L}(d | \vec{\theta}, \mathcal{H} )$. Under the hypothesis that the data, $d$, consists of Gaussian, stationary noise $n$ and a gravitational-wave signal $h(\vec{\theta})$, the likelihood is a Gaussian \cite{VV:2010}:
\beq
\mathcal{L}(d| \vec{\theta}, \mathcal{H} ) \propto e^{-(d - h(\vec{\theta})|d - h(\vec{\theta}) )/2}, \
\label{eq:likelihood}
\eeq
where $(a|b)$ is the usual noise-weighted inner-product,
\beq
(a|b) = 4\Re\int^{f_{max}}_{f_{min}}df \ \frac{\tilde{a}(f)\tilde{b}^{*}(f)}{S_{n}(f)},\
\eeq
and $S_{n}(f)$ is the detector's noise power spectral density (PSD). The limits of integration correspond to the bandwidth of the detector. A significant computational cost of evaluating the likelihood comes from computing the template waveform $h(\vec{\theta})$ at each point in the parameter space. 

The full PDF is multi-dimensional and to get estimates on individual parameters we work with the \textit{marginalized} PDF of either a single parmeter $\theta_{A} \in \vec{\theta}$ or pairs of parameters $(\theta_{A}, \theta_{B}) \in \vec{\theta}$. Writing $\vec{\theta} = (\theta_{A}, \vec{\Theta})$, the marginalized one-dimensional PDF of $\theta_{A}$ is thus:
\beq
p(\theta_{A}| d, \mathcal{H}) = \int_{\vec{\Theta}}\ d\vec{\Theta}\ p(\vec{\theta}| d, \mathcal{H}) ,\
\label{eq:marginalized}
 \eeq
and the extension to pairs is trivial.

To efficiently evaluate the likelihood function we typically use a stochastic sampling algorithm. Here we employ Markov-chain Monte Carlo (MCMC), whose application to gravitational-wave parameter estimation is described in \cite{VDS:2008}. 
The basic element of a MCMC is the ``Markov chain'' which directly samples the posterior distribution, after an initial ``burn-in'' is discarded.

We use the stationary phase approximation (SPA) inspiral waveforms for both our simulated signal and template model.  This allows us to directly ``inject'' the signal waveform into simulated frequency-domain noise without performing an additional Fourier transformation, which could introduce spurious artifacts related to the abrupt in-band termination of the time-domain waveform.  For fixed $\alpha$, $\delta$, $\iota$, and $\varphi$, the post-Newtonian frequency-domain waveform has the form
\beq
\tilde{h}(f) = \frac{\mathcal{A}(\mathcal{M}, \eta; f)}{D} e^{2\pi i f t_c - i \phi_c + i \Psi(\mathcal{M}, \eta; f)},
\label{eq:spa}
\eeq
where expressions for the amplitude $\mathcal{A}(\mathcal{M}, \eta; f)$ and phase $\Psi(\mathcal{M}, \eta; f)$ can be found in \cite{Buonanno:2009} for TaylorF2 post-Newtonian waveforms at zeroth order in amplitude and 3.5 post-Newtonian order in phase.

In general, the template model will be determined by the needs of a particular analysis, however for this proof of principle we will not consider time-domain template waveforms.

\section{Interpolating template waveforms using the SVD} 
\label{sec:interp_svd}

The interpolation scheme used in \cite{Cannon:2012} is based on the truncated SVD of a gravitational-wave template bank. The SVD decomposes the template bank into a set of orthogonal basis templates, the number of which equals the number of templates in the bank, and projection coefficients. Any template in the bank can be reconstructed from the bases weighted by the appropriate projection coefficients. However not all the bases are required to approximately reconstruct the waveforms. Truncating the SVD reduces the number of unique basis templates; we choose to truncate so that the norm of any reconstructed template is conserved to a level of $\sim 10^{-5}$ (c.f.~Eqs.~(27) and (28) in \cite{Cannon:2010}). The coefficients can be interpolated within the domain of the template bank which takes us from a discrete description of the template bank to a continuous one. Any template waveform in this domain is then approximately recovered by a linear combination of the basis templates weighted by the appropriate \textit{interpolated} projection coefficients.

Specifically, the SVD of a template bank of gravitational waveforms allows them to be written as a linear combination of basis waveforms $\vec{u}^{\mu}$ with projection coefficients $M_{\mu}(\mathcal{M}_{k}, \eta_{l})$ where the indices $k$ and $l$ enumerate a particular template in the bank. This implicitly assumes that the template bank follows a rectangular grid in $(\mathcal{M}, \eta)$. The waveform can thus be written
\beq
\vec{h}(\mathcal{M}_{k}, \eta_{l}) = \sum_{\mu}M_{\mu}(\mathcal{M}_{k}, \eta_{l})\vec{u}^{\mu}.\
\label{eq:SVD_template}
\eeq

To find the basis vectors $\vec{u}^{\mu}$ we compute the SVD of a real-valued $N\times L$ matrix, with $N \leq L$, \textbf{H} whose rows are template waveforms. The SVD factors \textbf{H} as
\beq
H_{\nu j} = \sum_{\mu =1}^{N}v_{\nu \mu}\sigma_{\mu}u_{\mu j},\
\eeq
where \textbf{v} is a unitary matrix of reconstruction coefficients, $\vec{\sigma}$ is a vector of singular values and \textbf{u} is a matrix of orthonormal basis vectors $\vec{u}$. In the above formalism the projection coefficients are related to the reconstruction-coefficient matrix and vector of singular values by $M_{\mu}(\mathcal{M}_{k}, \eta_{l}) := \sigma_{\mu}(v_{(2\nu)\mu} + iv_{(2\nu+1)\mu})$ where $\nu$ enumerates a (complex) waveform in the template bank. The indices $(k, l)$ are related to $\nu$ by the specific packing of the matrix \textbf{H}, the details of which are irrelevant.

The projection coefficients can be interpolated and we employ the method of \cite{Cannon:2012}, using Chebyshev polynomials of the first kind (c.f.~Eqs.~(7) and (8) in \cite{Cannon:2012}).  An ``interpolated waveform'', $\vec{h}^{\prime}$, can thus be constructed according to \eqref{eq:SVD_template} from a linear combination of interpolated coefficients and basis waveforms:

\beq
\vec{h}^{\prime}(\mathcal{M}, \eta) = \sum_{\mu}M^{\prime}_{\mu}(\mathcal{M}, \eta)\vec{u}^{\mu}.\
\label{eq:interp}
\eeq
This representation of $\vec{h}^{\prime}$ is continuous and hence any waveform within the domain of the original template bank can be recovered by interpolation. In general the accuracy of interpolated waveforms depends directly on the density of the template bank within a particular domain. Below we illustrate the application of the SVD to parameter estimation.

\section{Parameter Estimation using interpolated waveforms}
\label{sec:param_est}

We will compare the marginalized one-dimensional PDFs obtained by using an interpolated template waveform family to those generated using the standard, non-interpolated template waveform family for the same data set. For illustration we consider a toy example with five free parameters. We generate a single-detector data set $\vec{d}$ containing a signal waveform $h(\vec{\theta})$ and Gaussian stationary noise $n$ with a power spectral density roughly matching that of initial LIGO \cite{iLIGO}. By only having five free parameters we effectively set the prior on the other four to be delta functions centered on the signal values.  We chose to fix the sky position ($\alpha, \delta$) and the inclination and polarization angle ($\iota, \varphi$) of the template waveforms such that they are not searched over.   

Because interpolation is carried out in the mass space only, we focus on exploring the effects of interpolation on mass parameters and parameters that are known to be very strongly correlated with masses (time and phase of coalescence and distance).    If the accuracy of the recovery of these parameters is unaffected by interpolation, we do not expect the angular parameters to be affected, either.  However, it is important to realize that the absolute accuracy with which some parameters, particularly distance, are estimated is improved by fixing sky location and orientation parameters and lifting corresponding degeneracies.   Thus, the measurement uncertainties inferred below should not be considered typical.  Since we demand that systematic biases from using interpolated templates are smaller than statistical measurement uncertainties, the improvement in the accuracy of distance measurement means that we are being conservative in evaluating the quality of SVD-interpolated parameter estimation.

The signal contained in the data set has source parameters $(\mathcal{M},\ \eta,\ D,\ t_{c},\ \phi_{c}) = (\unit{7.45}{\smass},\ 0.247,\ \unit{33}{\Mparsec},\ \unit{0}{\second},\ 2.16)$ and we have omitted the others which are not searched over. The signal has a signal-to-noise ratio $\mathrm{SNR} = 14.8$. The frequency-domain data is sampled at $\Delta f =\unit{1/32}{\hertz}$ with a maximum frequency of $\unit{512}{\hertz}$. 

The prior distributions are set as follows. We use a uniform prior on $\log D$ and $\eta$ with ranges $D\in \left[\unit{1}{\Mparsec}, \unit{100}{\Mparsec}\right]$ and $\eta \in \left[0.175,\ 0.250\right]$. We use a prior on $\mathcal{M}$ of the form $\mathcal{P}(\mathcal{M}|\mathcal{H}) \propto \mathcal{M}^{-11/6}$ in the range $\mathcal{M} \in \left[\unit{7.20}{\smass},\ \unit{7.60}{\smass}\right]$. We use flat priors on $\phi_{c}$ and $t_{c}$ over the range $0 \leq \phi_{0} \leq 2\pi$ and $ \unit{-0.1}{\second} \leq t_c \leq \unit{0.1}{\second}$, respectively. The prior on $\mathcal{M}$ corresponds to the Jeffreys prior for the waveform family described by \eqref{eq:spa}  \cite{VV:2010}.

For the mock data set we ran a MCMC comprising five parallel Markov chains in order to compute the PDF $p(\vec{\theta} = (\mathcal{M},\ \eta,\ D,\ t_{c},\ \phi_{c})| d, \mathcal{H})$. The limits of integration of the likelihood function, \eqref{eq:likelihood}, are fixed to $f_{\mathrm{min}} = \unit{40}{\hertz}$ Hz, $f_{\mathrm{max}} = \unit{512}{\hertz}$. To extract the posterior samples from the raw MCMC output we discard the first $10,000$ samples as burn-in. 

We measure the convergence of the parallel chains using the Gelman-Rubin R-statistic \cite{GelmanRubin}. For well converged chains this should be close to $R=1$ and we regard the MCMC to be complete once $R \leq 1.001$ for all parameters. 

\subsection*{SVD Setup}

\label{sec:svd}

The input to the SVD is a set of whitened time domain waveforms \cite{Cannon:2012}. The frequency-domain SPA waveforms are whitened in the frequency domain with the PSD and transformed into the time-domain for interpolation. By carrying out the interpolation in the time domain, we show that the technique can be applied to time-domain waveform families, though in this example we could work directly in the frequency-domain. Time-domain waveforms are typically computationally expensive for parameter estimation (see Sec.~\ref{sec:comp_cost}), so this approach allows us to assess the computational savings associated with interpolating them. It is also consistent with the work in \cite{Cannon:2010,Cannon:2012}, where time-domain waveforms were interpolated.

We ensure that all  templates are of the same length, equal to the next highest power of two of the longest time-domain waveform in the set, which in our case is $\unit{2}{\second}$. For the proof of principle we apply the SVD to a small patch in $\mathcal{M}-\eta$ space bounding the signal parameters. This region is set by the prior range on $\mathcal{M}$ and $\eta$ described above, chosen to be broad enough so that, for our data set, there is no likelihood support near the boundaries.

\begin{figure}[htp]
\centering
\includegraphics[scale=0.35]{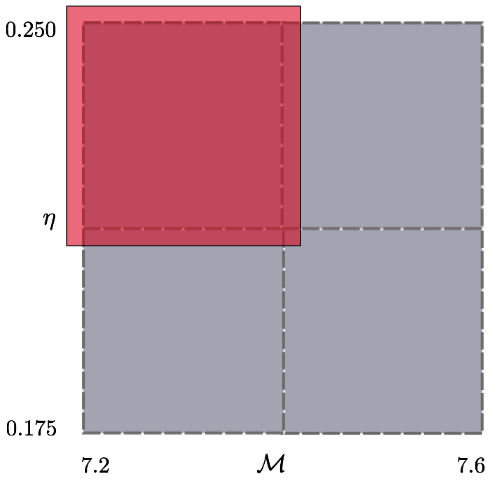}
\caption{Patching of the parameter space. The space is patched into four regions.  The SVD is applied to each patch (shown in red) which is padded at the boundary by one waveform and so contains a (15+1)$\times$(15+1) grid of template waveforms.  The SVD of the template banks consists of 20 basis templates in each patch.}
\label{fig:patches}
\end{figure}

The number of computations for the SVD of a $N\times L$ matrix with $N \leq L$ scales like $\mathcal{O}(LN^{2})$.  For the purposes of constructing the SVD we have found it efficient to split the $\mathcal{M}-\eta$ space into four equally sized patches, with a separate SVD applied to each patch. The density of waveforms in each patches' bank is chosen such that the normalized inner-product between non-interpolated waveforms and interpolated waveforms generated on an evenly spaced grid in each patch is at least $99.9\%$. Such normalized mismatches of $<0.001$ between interpolated and non-interpolated waveforms should ensure that parameter-estimation accuracy is not compromised as long as the the signal-to-noise ratio does not exceed $\sim 20$ (so that twice the mismatch times the square of the SNR is less than unity \cite{Lindblom:2008, Ohme:2012}), although parameter estimation could remain accurate at even higher SNRs.   For the mass space in this example, we find that a (15+1)$\times$(15+1) grid of template waveforms in each patch is sufficient for the required accuracy. We add one to the grid in each dimension because the interpolation can be unpredictable at the boundaries of the space. The patching is shown in Fig.~\ref{fig:patches}. Each waveform in the template bank is generated at a fiducial distance of 1 Mpc. The truncated-SVD of the template bank in each patch uses 20 basis waveforms. 

One subtlety of the implementation is that the interpolated waveforms are whitened time-domain filters. The likelihood function \eqref{eq:likelihood} is computed using un-whitened frequency series which in our example correspond to the Fourier transform of a complex time series. To recover the appropriate frequency series we first have to de-whiten the interpolated waveform by Fourier transforming the interpolated waveform and multiplying by the inverse of the PSD. While this process of de-whitening does not affect the normalized inner-product between interpolated and non-interpolated waveforms evaluated at the same point in parameter space, we have made no attempt to ensure the accuracy of the overall amplitude. We note that this is likely to have an impact on distance estimates. 

We further note that we do not use normalized waveforms as input to the SVD as was done in \cite{Cannon:2010, Cannon:2012}. Over large areas of the mass-space it is necessary to normalize the waveforms because there may be significant differences in the power of waveforms across the space and so the SVD can disproportionally weight waveforms with more power if they are not normalized. We are not affected by this issue because all the waveforms in our example have roughly the same power due to the limited extent of the template bank in mass space. In general one would use normalized waveforms for SVD over a larger parameter space; however, to recover the overall amplitude of the interpolated waveforms the normalization coefficients would themselves have to be interpolated. 

Below we compare the results of parameter estimation using interpolated and non-interpolated waveforms.

\section{Results: Comparison of parameter estimates using interpolated and non-interpolated waveforms}
\label{sec:results}

\begin{figure*}[htp]
\centering
\includegraphics[scale=0.64]{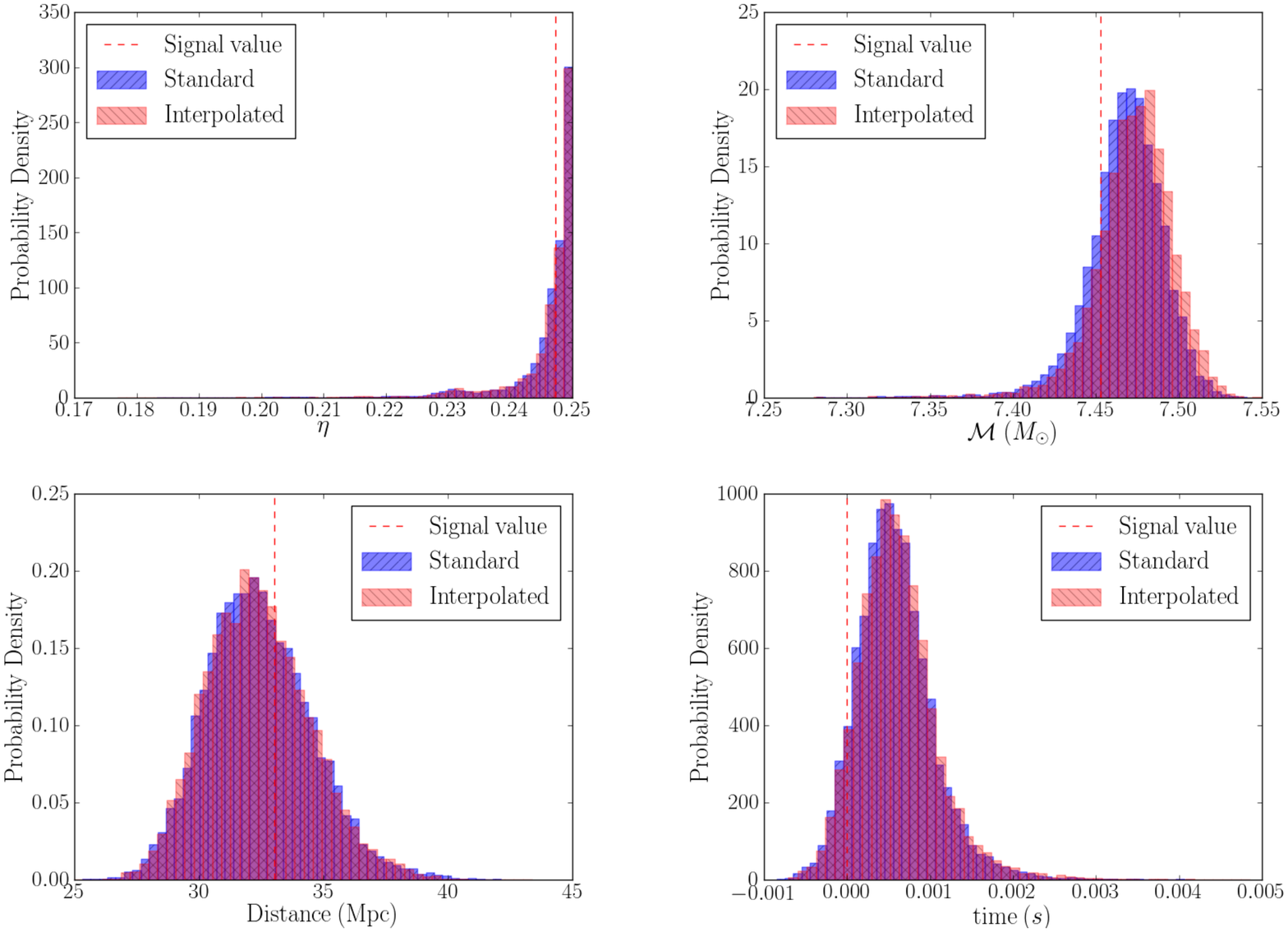}
\caption{Marginalized PDFs, \eqref{eq:marginalized}, produced using non-interpolated waveforms (blue or dark grey) and interpolated waveforms (red or light grey). The signal value is shown as the dashed red vertical line.}
\label{fig:pdfs_comps}
\end{figure*}

The marginalized PDFs for complete MCMC runs using non-interpolated and interpolated waveforms are shown in Fig.~\ref{fig:pdfs_comps}. We have omitted the marginalized one-dimensional PDF of the coalescence phase $\phi_{c}$ as it of little physical interest. Each run required around $1.5\times 10^{6}$ waveform computations. The mean posterior values of the distributions along with the signal values are shown in Table~\ref{table:summary_stats}. 

\begin{table}[!htp]
\begin{center}
    \begin{tabular}{ | l | p{2.5cm} | p{2.5cm} | l |}
    \hline
    Param & Mean posterior value (interpolated SPA) & Mean posterior value (SPA) & Signal Value \\ \hline
    $\mathcal{M}\ (M_{\odot}) $ & 7.472 (2.5$\times 10^{-2}$) & 7.467 (2.5$\times 10^{-2}$) & 7.450 \\ \hline
    $\eta$ & 0.2457 (7.1$\times 10^{-3}$) & 0.2457 (7.2$\times 10^{-3}$) & 0.2473 \\ \hline
    D (Mpc)& 32.39 (2.11) & 32.40 (2.13) & 33.00 \\ \hline
    $t_{c}\ (s)$ & $1.0\times10^{-3}$ (4$\times 10^{-4}$) & $1.0\times10^{-3}$ (4$\times 10^{-4}$) & 0 \\
    \hline
    \end{tabular}
    \caption{Maximum likelihood parameter estimates (and standard deviations) of the marginalized PDFs using interpolated and non-interpolated waveforms (Fig.~\ref{fig:pdfs_comps}).}
    \label{table:summary_stats}
\end{center}
\end{table}

The chirp-mass distribution computed using interpolated waveforms is clearly biased. This is corroborated by a two-sample K-S test which reveals that the two sets of samples are not consistent with arising from the same distribution with overwhelming odds. Nevertheless, the systematic bias in the chirp mass is a factor of four smaller than the statistical measurement uncertainty.  Thus, we pass a commonly-used threshold for sufficient waveform-model accuracy \cite[e.g.,][]{Ohme:2012}.  We note that the accuracy could be increased by, for example, using a higher density template bank or using normalized waveforms as input to the SVD. In general the required accuracy can be estimated from the detection trigger SNR \cite{Ohme:2012}. 

The two-sample K-S test marginally fails for the coalescence-time distribution, but there is no evidence of a systematic bias on the scale of statistical measurement errors.  We find that the sets of posterior samples for the other two PDFs in Fig.~\ref{fig:pdfs_comps}, symmetric mass ratio and distance, are consistent with arising from same distribution as quantified by the K-S test.

\subsection*{Computational cost of template waveform generation}

\label{sec:comp_cost}

Two commonly-used time-domain waveform families which are relevant for parameter estimation are the inspiral-only post-Newtonian approximant TaylorT4 \cite{Buonanno:2009} and the effective-one-body family calibrated to numerical relativity (EOBNR, \cite{Buonanno:2009}) that includes inspiral, merger, and ringdown phases. The latter are typically more computationally intensive.

For technical reasons, our comparison uses interpolated waveforms which are based on the SPA approximation rather than time-domain waveform families. This is inconsequential because the SVD procedure is the same regardless of the form of the input waveforms. We have observed that applying the truncated SVD to TaylorT4 time-domain templates in the mass space used to construct the SVD in Sec.~\ref{sec:svd} yields the same number of basis vectors as when applying it to inverse-FFT'd SPA templates. The computational cost of interpolation will therefore be identical.

Our measure to compare the computational costs of interpolated, TaylorT4 and EOBNR waveforms is the time it takes to compute a single interpolated waveform. While this does not estimate the theoretical minimum number of FLOPs of the process, and is also hardware dependent, it does provide a useful heuristic for comparing the relative speed of each waveform family. Recall that the interpolated waveforms are a time-domain approximant and hence the comparison is to determine the computational savings for time-domain waveforms. We restrict our comparison to waveforms generated in the mass space in Fig.~\ref{fig:patches}. The length of TaylorT4 and EOBNR waveforms will in general depend on the specific source masses. For a fair comparison we compare waveforms which have approximately the same number of data points. Because EOBNR must be generated at a sampling rate of \unit{4096}{\hertz}, we ensure that the interpolated and TaylorT4 waveforms are sampled at this frequency. Each waveform is approximately \unit{2}{\second} in duration.

The results of the comparison are shown in Table ~\ref{table:comp_time}. For reference we also show the computational time of standard SPA waveforms. We find that on average, the interpolated waveforms are ten times faster to generate than TaylorT4 and fifteen times faster than EOBNR, a significant increase in computational efficiency. However, for the waveform parameters considered here, inspiral-only waveforms could be generated at lower sampling rates than the \unit{4096}{\hertz} required for EOBNR waveforms; therefore, the cost of constructing interpolated or TaylorT4 waveforms can be around four times smaller relative to EOBNR than the values quoted in Table ~\ref{table:comp_time}.

\begin{table}[!htp]
\begin{center}
    \begin{tabular}{ | l | l | l |}
    \hline
    Waveform Family & Computational Time ($T$) \\ \hline
	SPA & 0.2 \\ \hline
    Interpolated & 1 \\ \hline
    TaylorT4 & 10 \\ \hline
    EOBNR & 15 \\
    \hline
    \end{tabular}
    \caption{Computational time of template waveform generation in units of computational time of interpolated waveforms, T. EOBNR, TaylorT4 and interpolated waveform families are generated at a sampling rate of \unit{4096}{\hertz} and have a duration of $\unit{2}{\second}$. The interpolated waveforms consist of 20 pre-computed basis vectors. SPA waveforms are generated in the frequency domain; to ensure the SPA waveforms contain the same number of sample points they are generated at a sampling frequency $\Delta f = \unit{1/2}{\hertz}$ and have a maximum frequency of \unit{2048}{\hertz}.}
    \label{table:comp_time}
\end{center}
\end{table}
 
We also estimate the cost of pre-computing the SVD interpolation. We have previously noted that the computational cost of an SVD of an $N\times L$ matrix with $(N \leq L)$ scales like $\mathcal{O}(N^{2}L)$. One also needs to compute the $N\times L$ matrix of template waveforms as input to the SVD. The cost of computing a waveform of length $L$ is typically $\mathcal{O}(L)$, possibly with a very large pre-factor.  Thus, the total cost of pre-computing interpolation coefficients will be less than $\mathcal{O}(N^{2})$ times the cost of an individual waveform computation.  For instance, in our example, $N=16\times 16 = 256$, so interpolation can reduce overall MCMC costs for any time-domain waveform templates by an order of magnitude or more when the typical MCMC chain length of $\gtrsim 10^{6}$ samples is taken into account.

\section{Conclusion and Discussion}
\label{sec:conc}
We have provided a proof of principle that interpolated waveforms can be used for parameter estimation without unacceptable loss of accuracy. Our example was restricted to a five-dimensional search over the source chirp mass $\mathcal{M}$ and symmetric mass ratio $\eta$, the distance to the source $D$ and the time and phase at coalescence $t_{c}$ and $\phi_{c}$. We further restricted the prior ranges on $\mathcal{M}$ and $\eta$ to $\mathcal{M} \in \left[\unit{7.20}{\smass}, \unit{7.60}{\smass}\right]$ and $\eta \in \left[0.175,\ 0.250\right]$, respectively.  The systematic biases observed when using interpolated waveforms are demonstrated to be smaller than statistical measurement uncertainties.  Thus, SVD-interpolated waveforms satisfy the stringent waveform-model accuracy criteria imposed by parameter-estimation requirements.

The relative computational times of generating interpolated waveforms and time-domain TaylorT4 and EOBNR waveforms are shown in Table~\ref{table:comp_time}. Interpolated waveforms can be generated at around an order of magnitude more cheaply than TaylorT4 or EOBNR. This suggests that the computational cost of parameter estimation can be significantly reduced by employing SVD-interpolated waveforms for likelihood computation when the latter is dominated by the cost of waveform generation.

In order for interpolated templates to be viable for parameter estimation pipelines we need to apply the SVD-interpolation technique to a significantly larger region of the CBC mass space than in the example considered here. Searches of gravitational waves from low-mass systems look for binaries with a maximum total mass of \unit{25}{\smass} and a minimum component mass of \unit{1}{\smass} \cite{LVC:lowmass} and high mass searches target binaries with total masses between \unit{25}{\smass} and \unit{100}{\smass} \cite{LVC:highmass}. To be able to apply our parameter estimation technique to triggers from such searches in a single step, without first determining the more limited mass region where there is significant likelihood support, we will need to efficiently patch the parameter space over a large mass range so that the computational cost of generating the SVD can be minimized. A necessary condition for setting up the SVD in all patches is that its computational cost, plus the cost of generating interpolated waveforms, must be less than the cost of performing the parameter estimation with non-interpolated waveforms. This will be the subject of future work.

Furthermore, we have to be able to extend the SVD to generic waveform families. Particularly interesting is the potential to extend the technique to EOBNR waveforms, which are currently expensive to generate, and waveform families which describe binaries with spinning components. The latter class of waveforms have an intrinsic parameter space with up to six more independent parameters (two spin vectors) and the current technique of interpolation within the intrinsic parameter space of waveforms may become costly in large-dimensional spaces. However, it is interesting to consider the potential to apply the technique to spin-aligned/anti-aligned waveforms \cite[e.g.,][]{Ajith:2011,Taracchini:2012} as this class of waveforms have only one extra parameter, the reduced spin of the binary. The analysis of data from advanced LIGO and Virgo detectors, which may have lower-frequency cutoffs close to $\unit{10}{\hertz}$ \cite{Harry:2010}, will require template waveforms that are several minutes in duration. This technique is likely to be highly relevant to parameter estimation in that context.

\acknowledgments

We would like to thank Tyson Littenberg for reading through a draft
of the manuscript and providing us with useful comments.
DK is supported from the Max Planck Gesellschaft.  KC is supported by the
National Science and Engineering Research Council of Canada.  RJES acknowledges
support from a Perimeter Institute visiting graduate fellowship. Research at
Perimeter Institute is supported through Industry Canada and by the
Province of Ontario through the Ministry of Research \& Innovation.  This
document has LIGO document number LIGO-P1200136.

\bibliographystyle{apsrev}
\bibliography{./bib_file.bib}

\end{document}